# A mathematical theory of citing

M.V. Simkin and V.P. Roychowdhury
*Department of Electrical Engineering, University of California, Los Angeles, CA 90095-1594*

Recently we proposed a model, in which when a scientist writes a manuscript, he picks up several random papers, cites them and also copies a fraction of their references. The model was stimulated by our finding that a majority of scientific citations are copied from the lists of references used in other papers. It accounted quantitatively for several properties of empirically observed distribution of citations. However, important features, such as power-law distributions of citations to papers *published during the same year* and the fact that the average rate of citing *decreases with aging* of a paper, were not accounted for by that model. Here we propose a modified model: when a scientist writes a manuscript, he picks up several random *recent* papers, cites them and also copies some of their references. The difference with the original model is the word *recent*. We solve the model using methods of the theory of branching processes, and find that it can explain the aforementioned features of citation distribution, which our original model couldn't account for. The model can also explain "*sleeping beauties in science*", i.e., papers that are little cited for a decade or so, and later "awake" and get a lot of citations. Although much can be understood from purely random models, we find that to obtain a good quantitative agreement with empirical citation data one must introduce *Darwinian fitness* parameter for the papers.

## I. Introduction

A theory of citing was long called for by information scholars (Cronin, 1981). From a mathematical perspective an advance was recently made with the formulation and solution of the *model of random-citing scientists*[1] (Simkin & Roychowdhury, 2005a). According to the model, when a scientist writes a manuscript he picks up several random papers, cites them, and also copies a fraction of their references.

The model was stimulated by the recursive literature search model (Vazquez, 2001) and justified by the fact that a majority of scientific citations are copied from the lists of references used in other papers (Simkin & Roychowdhury, 2003, 2005b). The model leads to the cumulative advantage (Price, 1976) (also known today as preferential attachment; Barabasi and Albert, 1999) process, so that the rate of citing a particular paper is proportional to the number of citations it has already received. In spite of its simplicity, the model appeared to account for several major properties of empirically observed distributions of citations.

A more involved analysis, however, reveals that certain subtleties of the citation distribution are not accounted for by the model. It is known, that the cumulative advantage process would lead to oldest papers being most highly cited (Günter et al, 1996; Barabasi and Albert, 1999; Krapivsky and Redner, 2001)[2]. In reality, the average citation rate decreases as the paper in question gets older (Price, 1965; Nakamoto, 1988; Glänzel & Schoepflin, 1994;

---
[1] Random-citing model is used not to ridicule the scientists, but because it can be exactly solved using available mathematical methods, while yielding a better match with data than any existing model. This is similar to the random-phase approximation in the theory of an electron gas. Of course, the latter did not arouse as much protest, as the model of random-citing scientists, - but this is only because electrons do not have a voice. What is an electron? - Just a green trace on the screen of an oscilloscope. Meanwhile, within itself, electron is very complex and is as inexhaustible as the universe. When an electron is annihilated in a lepton collider, the whole universe dies with it. And as for the random-phase approximation: Of course, it accounts for the experimental facts - but so does the model of random-citing scientists.

[2] Some of these references do not deal with citing, but with other social processes, which are modeled using the same mathematical tools. In the present paper we rephrase the results of such papers in terms of citations for simplicity.

Pollmann, 2000). The cumulative advantage process would also lead to an exponential distribution of citations to papers of the same age (Günter et al, 1996; Krapivsky and Redner, 2001). Empirically it was found that citations to papers published during the same year are distributed according to a power-law (see the ISI dataset in Fig. 1(a) in Redner, 1998).

In the present paper we propose the ***modified model of random-citing scientists***: when a scientist writes a manuscript, he picks up several random ***recent*** papers, cites them and also copies some of their references. The difference with the original model is the word ***recent***. We solve this model using methods of the theory of branching processes (Harris, 1963) (we review its relevant elements in Appendix A), and show that it explains both the power-law distribution of citations to papers published during the same year and literature aging. A somewhat similar model was recently proposed by Bentley, Hahn & Shennan (2004) in the context of patents citations. However those authors just used it to explain a power law in citation distribution (for what the usual cumulative advantage model will do) and did not address the topics we just mentioned.

## II. Branching citations

While working on a paper, a scientist reads current issues of scientific journals and selects from them the references to be cited in it. These references are of two sorts:
- *Fresh papers he had just read* – to embed his work in the context of current aspirations.
- *Older papers that are cited in the fresh papers* he had just read – to position his work in the context of previous achievements.

It is not a necessary condition for the validity of our model that the citations to old papers are copied, but the paper itself remains unread (although such opinion is supported by the studies of misprint propagation; Simkin & Roychowdhury, 2003, 2005b). The necessary conditions are as follows:
- Older papers are considered for possible citing only if they were recently cited.
- If a citation to an old paper is followed and the paper is formally read – the scientific qualities of that paper do not influence its chance of being cited[3].

A reasonable estimate for the length of time a scientist works on a particular paper is one year. We will thus assume that "recent" in the model of random-citing scientists means preceding year. To make the model mathematically tractable we enforce time-discretization with a unit of one year. The precise model to be studied is as follows. Every year $N$ papers are published. There is, on average, $N_{ref}$ references in a published paper (the actual value is somewhere between 20 and 40). Each year, a fraction $\alpha$ of references goes to randomly selected preceding year papers (the estimate[4] from actual citation data is $\alpha \approx 0.1$ (see Fig. 4 in Price, 1965) or $\alpha \approx 0.15$ (see Fig. 6 in Redner, 2004). The remaining citations are randomly copied from the lists of references used in the preceding year papers.

When $N$ is large, this model leads to the first-year citations being Poisson-distributed. The probability to get $n$ citations is

$$p(n) = \frac{\lambda_0^n}{n!} e^{-\lambda_0}, \qquad (1)$$

were $\lambda_0$ is the average expected number of citations

$$\lambda_0 = \alpha N_{ref}. \qquad (2)$$

The number of the second-year citations, generated by each first year citation (as well as, third-year citations generated by each second year citation and

---

[3] This assumption may seem radical, but look at the following example. The writings of J. Lacan (10,000 citations) and G. Deleuze (8,000 citations) were argued to be nonsense (Sokal & Bricmont (1998). Sadly enough, the work of the true scientists is far less cited: A. Sokal – 2,700 citations, J. Bricmont – 1,000 citations.

[4] The uncertainty in the value of $\alpha$ depends not only on the accuracy of the estimate of the fraction of citations which goes to previous year papers. We also arbitrarily defined recent paper (in the sense of our model), as the one published within a year. Of course, this is by order of magnitude correct but the true value can be anywhere between half a year and two years.

so on), again follows a Poisson distribution, this time with the mean

$$\lambda = (1-\alpha). \quad (3)$$

Within this model, citation process is a branching process (see Appendix A) with the first year citations equivalent to children, the second-year citations to grand children, and so on.

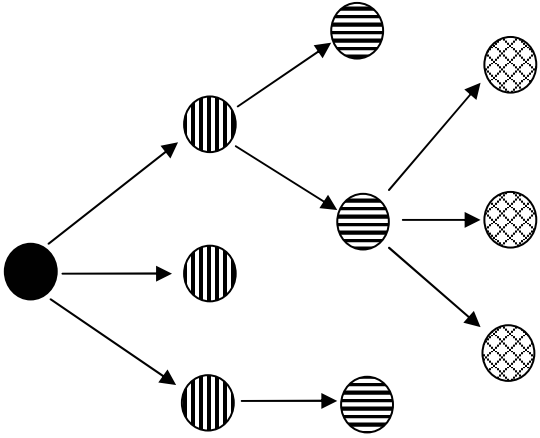

**Figure 1.** An illustration of the branching citation process, generated by the modified model of random-citing scientists. During the first year after publication, the paper was cited in three other papers written by the scientists who have read it. During the second year one of those citations was copied in two papers, one in a single paper and one was never copied. This resulted in three second year citations. During the third year, two of these citations were never copied, and one was copied in three papers.

As $\lambda < 1$, this branching process is subcritical. Figure 1 shows a graphical illustration of the branching citation process.

Substituting Eq. 1 into Eq. A1 we obtain the generating function for the first year citations:

$$f_0(z) = e^{(z-1)\lambda_0}. \quad (4a)$$

Similarly, the generating function for the later-years citations is:

$$f(z) = e^{(z-1)\lambda}. \quad (4b)$$

The process is easier to analyze when $\lambda = \lambda_0$, or $\frac{\lambda_0}{\lambda} = \frac{\alpha}{1-\alpha} N_{ref} = 1$, as then we have a simple branching process, where all generations are governed by the same offspring probabilities (the case when $\lambda \neq \lambda_0$ is studied in Appendix B).

## A. Distribution of citations to papers *published* during the same year

Theory of branching processes allows us to analytically compute the probability distribution, $P(n)$, of the total number of citations the paper receives before it is finally forgotten. This should approximate distribution of citations to old papers. Substituting Eq. 4b into Eq. A11 we get:

$$P(n) = \frac{1}{n!}\left[\frac{d^{n-1}}{d\omega^{n-1}}e^{n(\omega-1)\lambda}\right]_{\omega=0}$$
$$= \frac{(n\lambda)^{n-1}}{n!}e^{-\lambda n} \quad (5)$$

Applying Stirling's formula to Eq. 5, we get that large $n$ asymptotic of the distribution of citations is:

$$P(n) \propto \frac{e}{\lambda\sqrt{2\pi}}\frac{1}{n^{3/2}}e^{-(\lambda-1-\ln\lambda)n} \quad (6)$$

When $1-\lambda \ll 1$ the factor in the exponent can be approximated as:

$$\lambda - 1 - \ln\lambda \approx (1-\lambda)^2/2. \quad (7)$$

As $1-\lambda \ll 1$, the above number is small. This means that for $n \ll 2/(1-\lambda)^2$ the exponent in Eq. 6 is approximately equal to 1 and the behavior of $P(n)$ is dominated by the $\frac{1}{n^{3/2}}$ factor. In contrast, when $n \gg 2/(1-\lambda)^2$ the behavior of $P(n)$ is dominated by the exponential factor. Thus citation distribution changes from a power law to an exponential (suffers an exponential cut-off) at about

$$n_c = \frac{1}{\lambda - 1 - \ln\lambda} \approx \frac{2}{(1-\lambda)^2} \quad (8)$$

citations. For example, when $\alpha = 0.1$, Eq. 3 gives $\lambda = 0.9$ and from Eq. 8 we get that the exponential cut-off happens at about 200 citations. We see that the model is capable of a qualitative explanation of the power law distribution of citations to papers of the same age. The exponential cut-off at 200, however, happens too soon, as the actual citation distribution obeys a power law well into thousands of citations. In the following sections we will show that taking into account the effects of literature growth and of variations in papers' Darwinian fitness can fix this.

**B. Distribution of citations to papers *cited* during the same year**

In Appendix A we computed the fraction of families surviving after $k$ generations (Eq. A6), and their average sizes (Eq. A7). These results are directly applicable to the fraction of papers still cited $k$ years after publication, and the average number of citations, those papers receive during the $k^{th}$ year. Next we make an approximation assuming that all $k$-year old papers have the same number of citations, equal to the average given by Eq. A7. Then the number of citations depends on age only, and the number of cited papers of a given age is given by Eq. A6. After performing simple variables substitutions, and noting that in our case $f''(1) = \lambda^2$, we get that the citation probability distribution is:

$$p(n) \approx \begin{cases} \frac{2}{\lambda^2} \frac{1}{n} & \text{when} \quad n < \frac{\lambda^2}{2(1-\lambda)} \\ 0 & \text{when} \quad n > \frac{\lambda^2}{2(1-\lambda)} \end{cases}. \quad (9a)$$

We can obtain a more accurate approximation taking into account the fact that the distribution of the sizes of surviving families is exponential (see Chapt. IV in Fisher, 1958):

$$p(n) \approx \frac{2}{\lambda^2} \frac{1}{n} e^{-\frac{2(1-\lambda)}{\lambda^2}n}. \quad (9b)$$

A similar formula was previously derived (using a different method) by Kimura and Crow (1964) and by Ewens (1964) in the context of frequency distribution of selectively neutral alternative forms of a gene in a biological population. The model used in those papers is practically identical to ours with $\alpha$ being the mutation rate, instead of the fraction of citations going to new papers. The theory developed by Kimura and Crow (1964) and by Ewens (1964) was subsequently used to study cultural transmission and evolution (Cavalli-Sforza & Feldman, 1981). Recent examples include the modeling of the dynamics of popularity of baby names [Hahn & Bentley, 2003)], and of dog breeds [Herzog, Bentley, and Hahn, 2004)].

**III. Scientific Darwinism**

Now we proceed to investigate the model, where papers are not created equal, but each has a specific *Darwinian fitness*, which is a bibliometric measure of scientific fangs and claws that help the paper to fight for citations with its competitors. In bibliometrics literature a similar parameter is sometimes called *latent rate* of acquiring citations (Burrell, 2003). While this parameter can depend on factors other than the intrinsic quality of the paper, the fitness is the only channel through which the quality can enter our model. The fitness may have the following interpretation. When a scientist writes a manuscript he needs to include in it a certain number of references (typically between 20 and 40, depending on implicit rules adopted by a journal where the paper is to be submitted). He considers random scientific papers one by one for possible citation, and when he has collected the required number of citations, he stops. Every paper has specific probability to be selected for citation, once it was considered for citation. We will call this probability a Darwinian fitness of the paper. Defined in such way, fitness is bounded between 0 and 1.

In this model a paper with fitness $\phi$ will on average have

$$\lambda_0(\varphi) = \alpha N_{ref} \varphi / \langle \varphi \rangle_p \quad (10)$$

first-year citations. Here we have normalized the citing rate by the *average fitness of published papers*, $\langle \varphi \rangle_p$, to insure that the fraction of citations going to previous year papers remained $\alpha$. The fitness distribution of references is different from the fitness distribution of published papers, as papers with higher fitness are cited more often. This distribution assumes an asymptotic form $p_r(\varphi)$, which depends on the distribution of the fitness of published papers, $p_p(\varphi)$, and other parameters of the model.

During later years there will be on average

$$\lambda(\varphi) = (1-\alpha)\varphi/\langle\varphi\rangle_r \qquad (11)$$

next year citations per one current year citation for a paper with fitness $\phi$. Here, $\langle\varphi\rangle_r$ is the *average fitness of a reference*.

**A. Distribution of citations to old papers *published* during the same year**

The average number of citations that a paper with fitness $\phi$ acquires during its cited lifetime is:

$$N(\varphi) = \lambda_0(\varphi)\sum_{n=0}^{\infty}(\lambda(\varphi))^n = \frac{\lambda_0(\varphi)}{1-\lambda(\varphi)} = \alpha N_{ref}\frac{\varphi}{\langle\varphi\rangle_p}\frac{1}{1-(1-\alpha)\varphi/\langle\varphi\rangle_r} \qquad (12)$$

Obviously, $\langle\varphi\rangle_r$ is obtained self-consistently by averaging $\varphi N(\varphi)$ over $\phi$:

$$\langle\varphi\rangle_r = \frac{\int p_p(\varphi)\varphi N(\varphi)d\varphi}{\int p_p(\varphi)N(\varphi)d\varphi} \qquad (13)$$

Let us consider the simplest case when the fitness distribution, $p_p(\varphi)$, is uniform between 0 and 1. This choice is arbitrary, but we will see that the resulting distribution of citations is close to the empirically observed one. In this case, the average fitness of a published paper is, obviously $\langle\varphi\rangle_p = 0.5$. The average fitness of a reference is given by Eq. 13, which now becomes:

$$\langle\varphi\rangle_r = \frac{\int_0^1 \frac{\varphi^2 d\varphi}{1-\gamma\varphi/\langle\varphi\rangle_r}}{\int_0^1 \frac{\varphi d\varphi}{1-\gamma\varphi/\langle\varphi\rangle_r}}, \qquad (14)$$

where

$$\gamma = 1-\alpha.$$

After some transformations Eq. 14 reduces to:

$$\gamma - 1 = \frac{(\gamma/\langle\varphi\rangle_r)^2/2}{\ln(1-\gamma/\langle\varphi\rangle_r)+\gamma/\langle\varphi\rangle_r}. \qquad (15)$$

When $\gamma$ is close to 1, $\langle\varphi\rangle_r$ must be very close to $\gamma$, and we can replace it with the latter everywhere but in the logarithm to get:

$$\frac{\gamma}{\langle\varphi\rangle_r} = 1 - e^{-\frac{1}{2(1-\gamma)}-1} \qquad (16)$$

For papers of fitness, $\phi$, citation distribution is given by Eq. 5 (or Eq. 6) with $\lambda$ replaced with $\lambda(\varphi)$, given by Eq. 11:

$$P(n,\varphi) \propto \frac{e}{\lambda(\varphi)\sqrt{2\pi}}\frac{e^{-(\lambda(\varphi)-1-\ln\lambda(\varphi))n}}{n^{3/2}}. \qquad (17)$$

When $\alpha = 0.1$: we have $\gamma = 0.9$, and Eq. 16 gives $\gamma/\langle\varphi\rangle_r \approx 1-e^{-6}$. From Eq. 11 it follows that $\lambda(1) = \gamma/\langle\varphi\rangle_r$. Substituting this into Eq. 8 we get that the exponential cutoff for the fittest papers ($\varphi = 1$) starts at about 300,000 citations. In contrast, for the unfit papers the cut-off is even stronger than in the model without fitness. For example, for papers with fitness of $\varphi = 0.1$ [5] we get $\lambda(0.1) = 0.1\gamma/\langle\varphi\rangle_r \approx 0.1$ and the decay factor in the exponent becomes $\lambda(0.1)-1-\ln\lambda(0.1) \approx 2.4$. This cut-off is so strong than not even a trace of a power law distribution will remain for such papers.

To compute the overall probability distribution of citations we need to average Eq. 17 over fitness:

---

[5] In the biological case to get a Darwinian fitness of 0.1, one needs to have a major genetic disease like cystic fibrosis (see pp. 11-12 in Cavalli-Sforza & Feldman, 1981). In contrast, it seems that otherwise healthy people can be prolific producers of scientific writings with very low fitness.

$$P(n) \propto$$
$$\frac{e}{\sqrt{2\pi}} \frac{1}{n^{3/2}} \int_0^1 \frac{d\varphi}{\lambda(\varphi)} e^{-(\lambda(\varphi)-1-\ln\lambda(\varphi))n} \quad (18)$$

We will concentrate on the large $n$ asymptotic. Then only highest-fitness papers (which have $\lambda(\varphi)$ close to 1) are important and the integral in Eq. 18 can be approximated (using Eq. 8) as:

$$\int_0^1 d\varphi \exp\left(-\left[1-\varphi\frac{\gamma}{\langle\varphi\rangle_r}\right]^2 \frac{n}{2}\right) = \frac{\langle\varphi\rangle_r}{\gamma}\sqrt{\frac{2}{n}} \int_{\left(1-\frac{\gamma}{\langle\varphi\rangle_r}\right)\sqrt{\frac{n}{2}}}^{\sqrt{\frac{n}{2}}} dz\, e^{-z^2}$$

The upper limit in the above integral can be replaced with infinity when $n$ is large. The lower limit can be replaced with zero when $n \ll n_c$, where

$$n_c = 2\left(1-\frac{\gamma}{\langle\varphi\rangle_r}\right)^{-2}. \quad (19a)$$

In that case the integral is equal to $\sqrt{\pi}/2$, and Eq. 18 gives:

$$P(n) \propto \frac{e\langle\varphi\rangle_r}{2\gamma} \frac{1}{n^2}. \quad (19b)$$

In the opposite case, $n \gg n_c$, we get (Weisstein, b):

$$P(n) \propto \frac{e\langle\varphi\rangle_r}{4\gamma} \frac{\sqrt{n_c}}{n^{2.5}} e^{-\frac{n}{n_c}} \quad (19c)$$

When $\alpha = 0.1$ we have $\gamma = 0.9$, $1-\frac{\gamma}{\langle\varphi\rangle_r} \approx e^{-6}$, and $n_c = 3\times 10^5$.

Compared to the model without fitness, we have a modified power-law exponent (2 instead of 3/2) and a very much relaxed cut-off of this power law.

As was already mentioned, because of the uncertainty of the definition of "recent" papers, the exact value of $\alpha$ is not known. Therefore, we give $n_c$ for a range of values of $\alpha$ in Table 1. As long as $\alpha \leq 0.15$ the value of $n_c$ does not contradict to the existing citation data.

**Table 1.** The onset of exponential cut-off in the distribution of citations, $n_c$, as a function of $\alpha$, computed using Eq. 19a.

| $\alpha$ | 0.3 | 0.25 | 0.2 | 0.15 | 0.1 | 0.05 |
|---|---|---|---|---|---|---|
| $n_c$ | 167 | 409 | 1405 | 9286 | 3.1E+05 | 7.2E+09 |

The major results, obtained for the uniform distribution of fitness, also hold for a non-uniform distribution which approaches some finite value at its upper extreme $p_p(\varphi=1) = a > 0$. In the Appendix C we show that in this case $\gamma/\langle\varphi\rangle_r$ is very close to unity when $\alpha$ is small. Thus we can treat Eq. 18 the same way we did it in the case of the uniform distribution of fitness. The only change is that Eqs. 19b and 19c acquire a pre-factor of $a$.

Things turn out different when $p_p(\varphi=1) = 0$. In Appendix C we consider a fitness distribution, which vanishes at $\varphi = 1$ as a power law: $p_p(\varphi) = \frac{1}{\theta+1}(1-\varphi)^\theta$. When $\theta$ is small ($\theta < \frac{2\times\alpha}{1-\alpha}$) the behavior of the model is similar to what was in the case of a uniform fitness distribution. The distribution of the fitness of cited papers $p_r(\varphi)$ approaches some limiting form with $\gamma/\langle\varphi\rangle_r$ being very close to unity when $\alpha$ is small. The exponent of the power law is, however, no longer 2 as it was in the case of a uniform fitness distribution (Eq 19b), but $2+\theta$. Note, however that when $\theta > \frac{2\times\alpha}{1-\alpha}$ the model behaves completely different (see the end of Appendix C).

Thus, a wide class of fitness distributions produces citation distributions very similar to the experimentally observed one. More research is needed to infer the actual distribution of the Darwinian fitness of scientific papers. However, two things are clear:
- Some variance in fitness is needed to account for the empirical data.
- The fitness distribution does not need to have a heavy tail.

**B. Distribution of citations to papers *cited* during the same year**

This distribution in the case without fitness is given in Eq. 9b. To account for fitness we need to replace $\lambda$ with $\lambda(\varphi)$ in Eq. 9b and integrate it over $\varphi$. The result is:

$$p(n) \propto \frac{1}{n^2} e^{-n/n_c^*}, \qquad (20a)$$

where

$$n_c^* = \frac{1}{2}\left(1 - \frac{\gamma}{\langle\varphi\rangle_r}\right)^{-1}. \qquad (20b)$$

Note that $n_c^* \sim \sqrt{n_c}$. This means that the exponential cut-off starts much sooner for the distribution of citation to papers *cited* during the same year, then for citation distribution for papers *published* during the same year.

The above results qualitatively agree with the empirical data for papers cited in 1961 (see Fig.2 in Price, 1965). The exponent of the power law of citation distribution reported in that work is, however, between 2.5 and 3. Quantitative agreement thus may be lacking.

## IV. Effects of literature growth

Up to now we implicitly assumed that the yearly volume of published scientific literature does not change with time. In reality, however, it grows, and does so exponentially (Asimov (1958) gives a vivid account).To account for this, we introduce a Malthusian parameter, $\beta$, which is yearly percentage increase in the yearly number of published papers. From the data on the number of items in the Mathematical Reviews Database[6] we obtain that the literature growth between 1970 and 2000 is consistent with $\beta \approx 0.045$. From the data on the number of source publications in the ISI database (see Table 1 in Nakamoto, 1988) we get that the literature growth between 1973 and 1984 is characterized by $\beta \approx 0.03$. One can argue that the growth of the databases reflected not only growth of the volume of scientific literature, but also increase in activities of Mathematical Reviews and ISI and true $\beta$ must be less. One can counter-argue that may be ISI and Mathematical Reviews could not cope with literature growth and $\beta$ must be more. Another issue is that the average number of references in papers also grows. What is important for our modeling is the yearly increase not in number of papers, but in the number of citations these papers contain. Using the ISI data we get that this increase is characterized by $\beta \approx 0.05$. As we are not sure of the precise value of $\beta$, we will be giving quantitative results for a range of its values.

---

[6] Growth in the total number of items in the MR Database since its founding in 1940:
http://www.ams.org/publications/60ann/FactsandFigures.html

**A. Model without fitness**

At first, we will study the effect of $\beta$ in the model without fitness. Obviously, Equations 2 and 3 will change into:

$$\lambda_0 = \alpha(1+\beta)N_{ref}, \qquad (21a)$$
$$\lambda = (1-\alpha)(1+\beta). \qquad (21b)$$

The estimate of the actual value of $\lambda$ is: $\lambda \approx (1-0.1)(1+0.05) \approx 0.945$. Substituting this into Equation 8 we get that the exponential cut-off in citation distribution now happens after about 660 citations.

A curious observation is that when the volume of literature grows in time the average amount of citations a paper receives, $N_{cit}$, is bigger than the average amount of references in a paper, $N_{ref}$. Elementary calculation gives:

$$N_{cit} = \sum_{m=0}^{\infty} \lambda_0 \lambda^m = \frac{\lambda_0}{1-\lambda} = \frac{\alpha(1+\beta)N_{ref}}{1-(1-\alpha)(1+\beta)} \qquad (22)$$

As we see $N_{cit} = N_{ref}$ only when $\beta = 0$ and $N_{cit} > N_{ref}$ when $\beta > 0$. There is no contradiction here if we consider an infinite network of scientific papers, as one can show using methods of the set theory (Kleene, 1952) that there are one-to-many mappings of an infinite set on itself. When we consider real, i.e. finite, network where the number of citations is obviously equal to the number of references we recall that $N_{cit}$, as computed in Eq. 22, is the number of citations accumulated by a paper during its cited lifetime. So recent papers did not yet receive their share of citations and there is no contradiction again.

**B. Model with Darwinian fitness**

Taking into account literature growth leads to transformation of Equations 10 and 11 into:

$$\lambda_0(\varphi) = \alpha(1+\beta)N_{ref}\varphi/\langle\varphi\rangle_p, \quad (23a)$$
$$\lambda(\varphi) = (1-\alpha)(1+\beta)\varphi/\langle\varphi\rangle_r. \quad (23b)$$

As far as the average fitness of a reference, $\langle\varphi\rangle_r$, goes, $\beta$ has no effect. Clearly, its only result is to increase the number of citations to all papers (independent of their fitness) by a factor $1+\beta$. Therefore $\langle\varphi\rangle_r$ is still given by Eq. 15. While, $\lambda(\varphi)$ is always less then unity in the case with no literature growth, it is no longer so when we take this growth into account. *When $\beta$ is large enough, some papers can become supercritical.* The critical value of $\beta$, i.e. the value which makes papers with $\varphi = 1$ critical, can be obtained from Eq. 23b:

$$\beta_c = \langle\varphi\rangle_r/(1-\alpha) - 1. \quad (24)$$

When $\beta > \beta_c$, a finite fraction of papers becomes supercritical. The rate of citing them will increase with time. Note, however, that it will increase always slower than the amount of published literature. Therefore, the relative fraction of citations to those papers to the total number of citations will decrease with time.

Critical values of $\beta$ for several values of $\alpha$ are given in Table 2. For realistic values of parameters ($\alpha \leq 0.15$ and $\beta \geq 0.03$) we have $\beta > \beta_c$ and thus our model predicts the existence of supercritical papers. Note, however, that this conclusion also depends on the assumed distribution of fitness.

**Table 2** Critical value of the Malthusian parameter $\beta_c$ as a function of $\alpha$ computed using Eq. 24. When $\beta > \beta_c$ the fittest papers become supercritical.

| $\alpha$ | 0.3 | 0.25 | 0.2 | 0.15 | 0.1 | 0.05 |
|---|---|---|---|---|---|---|
| $\beta_c$ | 0.12 | 0.075 | 0.039 | 0.015 | 2.6E-03 | 1.7E-05 |

It is not clear whether supercritical papers exist in reality or are merely a pathological feature of the model. Supercritical papers probably do exist if one generalizes "citation" to include references to a concept, which originated from the paper in question. For instance, these days a negligible fraction of scientific papers which use Euler's Gamma function contain a reference to Euler's original paper. It is very likely that the number of papers mentioning Gamma function is increasing year after year.

Let us now estimate the fraction of supercritical papers predicted by the model. As $(1-\alpha)/\langle\varphi\rangle_r$ is very close to unity, it follows from Eq. 23b that papers with fitness $\varphi > \varphi_c \approx 1/(1+\beta) \approx 1-\beta$ are in the supercritical regime. As $\beta \approx 0.05$, about 5% of papers are in such regime. This does not mean that 5% of papers will be cited forever, because being in supercritical regime only means having extinction probability less than one. To compute this probability we substitute Equations 23b and 4b into Eq. A3 and get:

$$p_{ext}(\varphi) = \exp((1+\beta)\times\varphi\times(p_{ext}(\varphi)-1)).$$

It is convenient to rewrite the above equation in terms of survival probability:

$$1 - p_{surv}(\varphi) = \exp(-(1+\beta)\times\varphi\times p_{surv}(\varphi)).$$

As $\beta \ll 1$ the survival probability is small and we can expand the RHS of the above equation in powers of $p_{surv}$. We limit this expansion to terms up to $(p_{surv})^2$ and after solving the resulting equation get:

$$p_{surv}(\varphi) \approx 2\frac{\varphi - \frac{1}{1+\beta}}{(1+\beta)\varphi} \approx 2(\varphi - 1 + \beta).$$

The fraction of forever-cited papers is thus: $\int_{1-\beta}^{1} 2(\varphi-1+\beta)d\varphi = \beta^2$. For $\beta \approx 0.05$ this will be one in four hundred. By changing the fitness distribution $p_p(\varphi)$ from a uniform this fraction can be made much smaller.

## V. Numerical simulations

The analytical results are of limited use, as they are exact only for infinitely old papers. To see what happens with finitely old papers, one has to do numerical simulations. Figure 2 shows results from such simulations (with $\alpha = 0.1$, $\beta = 0.05$, and uniform between 0 and 1 fitness distribution), i.e., distributions of citations to papers published within a single year, 22 years after publication. Results are compared with actual citation data for Physical Review D papers published in 1975 (as of 1997). Prediction of the cumulative advantage (Price, 1976) (AKA preferential attachment; Barabasi and Albert, 1999) model is also shown. As we mentioned earlier, that model leads to exponential distribution of citations to papers of same age, and thus can not account for highly-skewed distribution empirically observed.

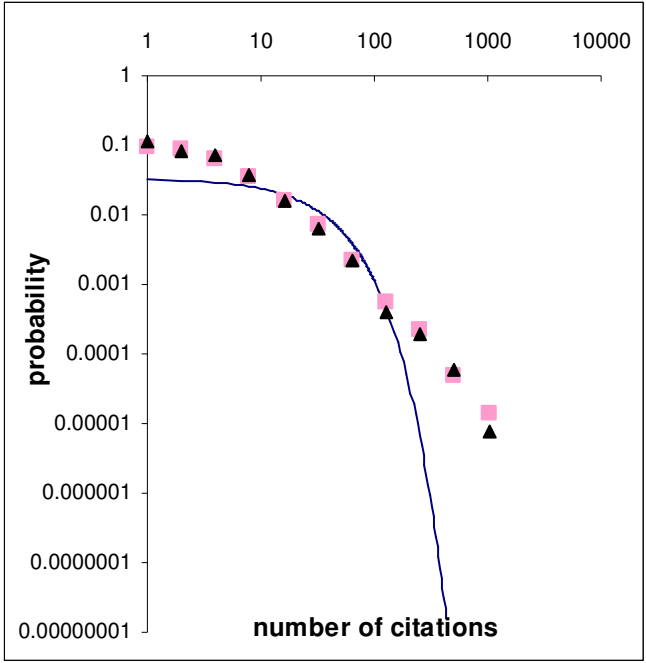

**Figure 2.** Numerical simulation of the *modified* model of random-citing scientists (triangles) compared to actual citation data for papers *published during a single year* (squares). The solid line is the prediction of cumulative advantage (AKA preferential attachment) model.

## VI. Unread citations

Recent scientific research points to the evidence that the majority of scientific citations were not read by the citing authors. Apart from the analysis of misprint propagation (Simkin & Roychowdhury, 2003, 2005b) this conclusion is indirectly supported by recent study (Brody and Harnad, 2005), which found that the correlation coefficient between the number of citations to and the number of readings of papers in arXiv.org is only $r$~0.45. This suggests that just 20% ($r^2$~0.2) of variance in number of citations is explained by the variance in the number of readings.

This should affect citation distribution in the model with fitness, because when paper is not read, its qualities can not affect its chance of being cited. Eq. 23a is obviously unchanged (since recent papers haven't yet been cited, citation could not be copied, so they had to be read). Equation 23b changes into:

$$\lambda(\varphi) = (1-\alpha)(1+\beta)(1 - R + R\varphi/\langle\varphi\rangle_r). \quad (25)$$

Here $R$ is the fraction of citations that are read by citing authors. It was estimated (Simkin & Roychowdhury, 2005b) to be $R = 0.2 \pm 0.1$.

Eq. 14 transforms into:

$$\langle\varphi\rangle_r = \frac{\int_0^1 \frac{\varphi^2 d\varphi}{1-\gamma(1-R+R\varphi/\langle\varphi\rangle_r)}}{\int_0^1 \frac{\varphi d\varphi}{1-\gamma(1-R+R\varphi/\langle\varphi\rangle_r)}}. \quad (26a)$$

After some transformations, Eq. 26a reduces to equation identical to Eq. 15 with $\gamma$ replaced with

$$\tilde{\gamma} = \frac{R\gamma}{1-\gamma(1-R)}. \quad (26b)$$

Approximation used in Eq. 16 is no longer valid as $\tilde{\gamma}$ is not close to one, and we have to solve Eq. 15 numerically.

A critical value of $\beta$ can be defined and computed similar to how it was done in Section IVb:

$$\beta_c = \frac{1}{1-\alpha}\frac{\langle\varphi\rangle_r}{(1-R)\langle\varphi\rangle_r + R} - 1. \quad (27)$$

Results are given in Table 3. We see that for realistic values of parameters ($\alpha \approx 0.15$, $\beta \approx 0.05$, and $R \approx 0.2$) we have $\beta < \beta_c$. That is, *unread citations can save us from supercritical papers*.

The argument in the beginning of this section, however, is not entirely correct. The fitness of a paper, apart from scientific qualities, which can only be assessed by reading, depends on scientific respectabilities of the associated authors, and of the journal where it was published. Besides, a paper's fitness may be reflected in the way it is referred to. So, perhaps, this section is only useful as a mathematical exercise.

**Table 3.** Critical value of the Malthusian parameter, $\beta_c$, as a function of $R$ for $\alpha = 0.15$, computed using Eq. 27.

| $R$ | 0 | 0.1 | 0.2 | 0.3 | 0.5 | 1 |
|---|---|---|---|---|---|---|
| $\beta_c$ | 0.18 | 0.13 | 0.10 | 0.08 | 0.05 | 0.015 |

## VII. Aging of scientific literature

Scientific papers tend to get less frequently cited as time passes since their publication. There are two ways to look at the age distribution of citations. One can take all papers *cited* during a particular year, and study the distribution of their ages. In Bibliometrics this is called *synchronous* distribution (Nakamoto, 1988). One can take all the papers *published* during a particular distant year, and study the distribution of the citations to these papers with regard to time difference between citation and publication. Synchronous distribution is steeper than the distribution of citation to papers published during the same year (see Figures 2 and 3 in Nakamoto, 1988). For example, if one looks at a synchronous distribution, than ten year old papers appear to be cited 3 times less than two year old papers. However, when one looks at the distribution of citations to papers published during the same year the number of citations ten years after publication is only 1.3 times less than two years after publication. The apparent discrepancy is resolved by noting that the number of published scientific papers had grown 2.3 times during eight years. When one plots not total number of citations to papers published in a given year, but the ratio of this number to the annual total of citations than resulting distribution (it is called *dyachronous* distribution (Nakamoto, 1988) is symmetrical to the synchronous distribution. Motylev (1989) used a similar procedure to refute the notion of more rapid aging of publications on rapidly developing fields of knowledge.

There is some controversy as of functional form of the citation age distributions. Nakamoto found it to be exponential for large ages. Pollmann (2000), however, states that a power law gives a better fit. Other proposed functional forms are lognormal, Weibull (AKA stretched or compressed exponential), and log-logistics distributions: see Burrell (2002) and references therein. Recently, Redner (2004) who analyzed a century worth of citation data from Physical Review had found that the synchronous distribution (he calls it citations *from*) is exponential, and the distribution of citations to papers published during the same year (he calls it citations *to*) is a power law with an exponent close to one. If one were to construct a dyachronous distribution using Redner's data, – it would be a product of a power law and an exponential function. Such distribution is difficult to tell from an exponential one. Thus, Redner's data may be consistent with synchronous and diachronous distributions being symmetric.

The predictions of the mathematical theory of citing are as follows. First, we consider the model without fitness. The average number of citations a paper receives during the $k^{th}$ year since its publication, $C_k$, is:

$$C_k = \lambda_0 \lambda^{k-1}, \qquad (28)$$

and thus, decreases exponentially with time. This is in qualitative agreement with Nakamoto's (1988) empirical finding. Note, however, that the exponential decay is empirically observed after the second year, with average number of the second year citations being higher than the first year. This can be understood as a mere consequence of the fact that it takes about a year for a submitted paper to get published.

Let us now investigate the effect of fitness on literature aging. Obviously, Eq. 28 will be replaced with:

$$C_k = \int_0^1 d\varphi \lambda_0(\varphi) \lambda^{k-1}(\varphi). \qquad (29)$$

Substituting Eqs. 10 and 11 into Eq. 29 and performing integration we get:

$$C_k = \frac{\alpha N_{ref}}{\langle\varphi\rangle_p}\left(\frac{\gamma}{\langle\varphi\rangle_r}\right)^{k-1}\frac{1}{k+1} \quad (30)$$

The average rate of citing decays with paper's age as a power law with an exponential cut-off. This is in agreement with Redner's data (See Fig.7 of Redner, 2004), though it contradicts the older work (Nakamoto, 1988), which found exponential decay of citing with time.

In our model, the transition from hyperbolic to exponential distribution occurs after about

$$k_c = -1/\ln(\gamma/\langle\varphi\rangle_r) \quad (31)$$

years. The values of $k_c$ for different values of $\alpha$ are given in Table 4. The values of $k_c$ for $\alpha \leq 0.2$ do not contradict the data reported by Redner (2004).

**Table 4.** The number of years, after which the decrease in average citing rate will change from a power law to an exponential, $k_c$, computed using Eq. 31, as a function of $\alpha$.

| $\alpha$ | 0.3 | 0.25 | 0.2 | 0.15 | 0.1 | 0.05 |
|---|---|---|---|---|---|---|
| $k_c$ | 9 | 14 | 26 | 68 | 392 | 59861 |

## VIII. Sleeping Beauties in science

Figure 3 shows two distinct citation histories. The paper, whose citation history is shown by the squares, is an ordinary paper. It merely followed some trend. When ten years later that trend got out of fashion the paper got forgotten. The paper, whose citation history is depicted by the triangles, reported an important but premature [Garfield, 1980; Glänzel and Garfield, 2004] discovery, significance of which was not immediately realized by scientific peers. Only ten years after its publication did the paper get recognition, and got cited widely and increasingly. Such papers are called "*Sleeping Beauties*" (Raan, 2004). Surely, the reader has realized that both citation histories are merely the outcomes of numerical simulations of the modified model of random-citing scientists.

After the original version of this paper was submitted for publication there appeared a paper by Burrell (2005) which used a phenomenological stochastic model of citation process to show that some sleeping beauties are to be expected by ordinary chance. An earlier paper (Glänzel, Schlemmer, & Thijs, 2003) ) addressed a similar issue using the cumulative advantage model. In this case, the authors were specifically concentrating on papers little cited during two years after publication (this is the standard time frame used in bibliometrics to determine the impact of a publication).

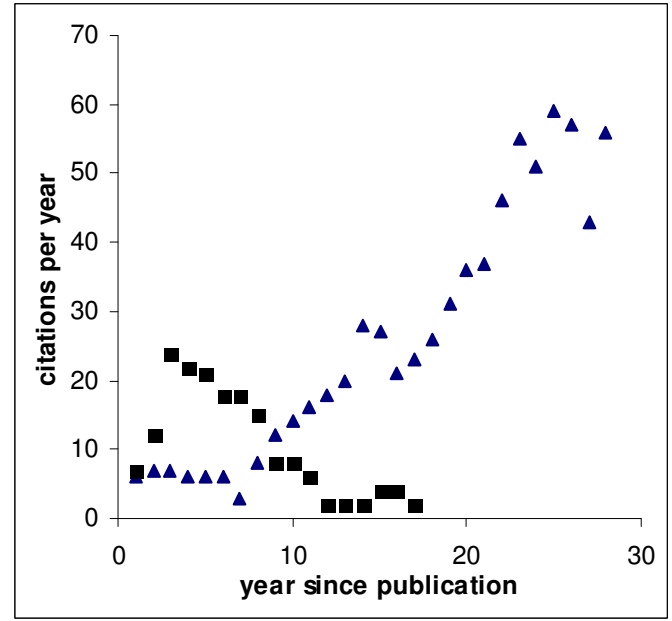

**Figure 3.** Two distinct citation histories: an ordinary paper (squares) and a "Sleeping Beauty" (triangles).

## IX. Relation to Self Organized Criticality

Those who are familiar with the Self Organized Criticality (SOC) of Bak, Tang, and Wiesenfeld (1988), may find of some interest that it is directly related to our study. We model scientific citing as a random branching process. In its mean-field version SOC can also be described as a branching process [Alstrøm, 1988; Lauritsen, Zapperi, and Stanley, 1996]. Here the sand grains, which are moved during the original toppling, are equivalent to sons. These displaced grains can cause further toppling, resulting in the motion of more grains, which are equivalent to grandsons, and so on. The total number of displaced grains is the size of the avalanche and is equivalent to the total offspring in the case of a branching process. Distribution of offspring is equivalent to distribution of avalanches in SOC.

Bak (1999) himself had emphasized the major role of chance in works of Nature: one sand grain falls, - nothing happens; another one (*identical*) falls, - and causes an avalanche. Applying these ideas to biological evolution, Bak & Sneppen (1993) argued that no cataclysmic external event was necessary to cause a mass extinction of dinosaurs. It could have been caused by one of many minor external events. Similarly, in the model of random-citing scientists: one paper goes unnoticed, but another one (*identical in merit*), causes an avalanche of citations. Therefore apart from explanations of 1/*f* noise, avalanches in sandpiles, and extinction of dinosaurs, the highly cited Science of Self Organized Criticality (Bak, 1999) can also account for its own success.

Now we would like to clarify some points of potential confusion.

Avalanches of citations, we are talking about, should not be confused with avalanches in power-law networks, which have been studied, for example, by Lee, Goh, Kahng, and Kim (2004). In the model of random citing scientists, the power law network of scientific papers *itself* is a product of avalanches.

Also note that the model of random citing scientists with Darwinian fitness is mathematically different from both the Bak-Sneppen (1993) model and from its modification by Vandewalle & Ausloos (1996). The model of random citing scientists reduces to a branching process, just like the aforementioned models. In addition in our model the fitness of new papers is uniformly distributed between 0 and 1, just like the fitness of the new species in the aforementioned models. However in our model the "offsprings" are citations, which carry the fitness of the cited paper. In the aforementioned models the "offsprings" are new species which are assigned a random fitness. As a result, our model leads to a different exponent of the avalanche distribution (2 instead of 1.5) than the mean-field versions of Bak-Sneppen and Vandewalle-Ausloos models.

## X. Conclusion

In the cumulative advantage (AKA preferential attachment) model, a power law distribution of citations is only achieved because papers have different ages. This is not immediately obvious from the early treatments of the problem [Simon, 1955; Price, 1976], but is explicit in later studies [Günter et al, 1996; Barabasi and Albert, 1999; Krapivsky and Redner, 2001]. In that model, the oldest papers are the most cited ones. The number of citations is mainly determined by a paper's age. At the same time, distribution of citations to papers of the same age is exponential [Günter et al, 1996; Krapivsky and Redner, 2001]. The key difference between that model and ours is as follows. In the cumulative advantage model, the rate of citation is proportional to the number of citations the paper had accumulated since its publication. In our model, the rate of citation is proportional to the number of citations the paper received during preceding year. This means that if an unlucky paper was not cited during previous year – it will never be cited in the future. This means that its rate of citation will be less than that in a cumulative advantage model. On the other hand, the lucky papers, which were cited during the previous year, will get all the citation share of the unlucky papers. Their citation rates will be higher than in the cumulative advantage model. There is thus more stratification in our model than in the cumulative advantage model. As a consequence, the resulting citation distribution is far more skewed.

One can argue that the cumulative advantage model with multiplicative fitness (Bianconi and Barabási, 2001) can explain a power-law distribution of citations to the same-year papers, when the distribution of fitness is exponential (see Appendix B in Simkin & Roychowdhury, 2006). Note, however, that this model is not capable of explaining literature aging.

This is the first paper that derives literature aging from a realistic model of scientist's referencing behavior. Stochastic models have been used previously to study literature aging, but they were of the phenomenological type. Glänzel and Schoepflin (1994) [7] used a modified cumulative advantage model, where the rate of citing is proportional to the product of the number of accumulated citations and some factor, which decays with age. Burrell (2003), who modeled citation process as a non-homogeneous Poisson process had to postulate some obsolescence distribution function.

---

[7] A very similar model was recently proposed by Dorogovtsev and Mendes (2002) (see Chapt. 9.7 ) in the context of growing networks.

In both these cases, aging was inserted by hand. In contrast, in our model, literature ages naturally.

## Appendix A: Theory of branching processes

This theory was conceived in the 19th century, when some British gentlemen noticed that many families that had occupied conspicuous positions in the past became extinct. At first they concluded that an increase in intellectual capacity is accompanied by a decrease in fertility. Afterward, the theory of branching processes was developed, which showed that a large proportion of families (or surnames) should become extinct by the ordinary law of chances.

Watson and Galton (1875) considered a model where in each generation, $p(0)$ per cent of the adult males have no sons, $p(1)$ have one son and so on. The problem is best tackled using the method of generating functions (Harris, 1963), which are defined as:

$$f(z) = \sum_{n=0}^{\infty} p(n) z^n . \tag{A1}$$

These functions have many useful properties, including that the generating function for the number of grandsons is $f_2(z) = f(f(z))$. To prove this notice that if we start with two individuals instead of one, and both of them have offspring probabilities described by $f(z)$, their combined offspring has generating function $(f(z))^2$. This can be verified by observing that the $n$th term in the expansion of $(f(z))^2$ is equal to $\sum_{m=0}^{n} p(n-m) p(m)$, which is indeed the probability that the combined offspring of two people is $n$. Similarly one can show that the generating function of combined offspring of $n$ people is $(f(z))^n$. The generating function for the number of grandsons is thus:

$$f_2(z) = \sum_{n=0}^{\infty} p(n)(f(z))^n = f(f(z)).$$

In a similar way one can show that the generating function for the number of grand-grandsons is $f_3(z) = f(f_2(z))$ and in general:

$$f_k(z) = f(f_{k-1}(z)). \tag{A2}$$

The probability of extinction, $p_{ext}$, can be computed using the self-consistency equation:

$$p_{ext} = \sum_{n=0}^{\infty} p(n) p_{ext}^n = f(p_{ext}). \tag{A3}$$

The fate of families depends on the average number of sons $\lambda = \sum n p(n) = [f'(z)]_{z=1}$. When $\lambda < 1$, Eq. A3 has only one solution, $p_{ext} = 1$, that is all families get eventually extinct (this is called subcritical branching process). When $\lambda > 1$, there is a solution where $p_{ext} < 1$, and only some of the families get extinct, while others continue to exist forever (this is called supercritical branching process). The intermediate case, $\lambda = 1$, is critical branching process, where all families get extinct, like in a subcritical process, though some of them only after very long time.

Though, for (sub) critical branching processes probability of extinction is unity, still a non-trivial quantity is the probability, $p_{ext}(k)$, of extinction after $k$ generations. Obviously:

$$p_{ext}(k) = f_k(0).$$

As $p_{ext} = 1$, than, for large $k$, $p_{ext}(k)$ must be close to 1. Therefore

$$f_k(0) = f(f_{k-1}(0)) \approx f(1) + f'(1)(f_{k-1}(0)-1) + \frac{f''(1)}{2}(f_{k-1}(0)-1)^2$$

After noting that $f(1) = 1$ and $f'(1) = \lambda$, and defining the survival probability $p_s(k) = 1 - p_{ext}(k)$, the above equation can be rewritten as:

$$\frac{p_s(k)}{p_s(k-1)} = \lambda - \frac{f''(1)}{2} p_s(k-1). \tag{A4}$$

Let us first consider the case $\lambda = 1$. Eq. A4 can than be approximated by the differential equation

$$\frac{dp_s(k)}{dk} = -\frac{f''(1)}{2}(p_s(k))^2,$$

which has a solution

$$p_s(k) = \frac{2}{f''(1)k}. \qquad (A5a)$$

In the case when $\lambda$ is substantially less than one the second term in the R.H.S. of Eq. A4 can be neglected and the equation can be easily solved:

$$p_s(n) \sim \lambda^k. \qquad (A5b)$$

In general Eq. A3 can be approximately solved to get:

$$p_s(k) \approx \frac{2}{f''(1)} \frac{\lambda^k \ln(1/\lambda)}{1 - \lambda^k}. \qquad (A6)$$

When $\lambda$ is very close to but less than one, Eq. A6 has an intermediate asymptotic of the form of Eq. A5a when $k < k_c$, where

$$k_c \approx \frac{1}{1-\lambda}.$$

When $k > k_c$, Eq. A6 approaches the form of Eq. A5b.

Next we estimate the average size $\bar{s}(k)$ of families still surviving after $k$ generations. As the expectation value of the offspring after $k$ generations is, obviously, $\lambda^k$, we have:

$$\bar{s}(k) = \frac{\lambda^k}{p_s(k)}. \qquad (A7)$$

After substituting Eq. A5a into Eq. A7 we see that for the critical branching process the average size of surviving family linearly increases with the number of passing generations: $\bar{s}(k) \approx \frac{f''(1)}{2} k$. By substituting Eq. A5b into Eq. A7 we get that for subcritical branching process after large number of generations the average size of a surviving family approaches the fixed value:

$$\bar{s}(\infty) \approx \frac{f''(1)}{2\ln(1/\lambda)} \approx \frac{f''(1)}{2(1-\lambda)}.$$

For a subcritical branching process we will also be interested in the probability distribution, $P(n)$, of total offspring, which is the sum of the numbers of sons, grandsons, grand-grandsons and so on (to be precise we include self in this sum just for mathematical convenience). We define the corresponding generating function (Otter, 1949):

$$g(z) = \sum_{n=1}^{\infty} P(n) z^n. \qquad (A8)$$

Using an obvious self-consistency condition (similar to the one in Eq. A3) we get:

$$zf(g) = g \qquad (A9)$$

Using Lagrange expansion[8] we obtain from Eq. A9:

$$g = \sum_{n=1}^{\infty} \frac{z^n}{n!} \left[ \frac{d^{n-1}}{d\omega^{n-1}} (f(\omega))^n \right]_{\omega=0}. \qquad (A10)$$

And using Eq. A8 we get:

$$P(n) = \frac{1}{n!} \left[ \frac{d^{n-1}}{d\omega^{n-1}} (f(\omega))^n \right]_{\omega=0}. \qquad (A11)$$

Theory of branching processes is useful in many applications, for example, in the study of nuclear chain reactions.

Nuclei of uranium can spontaneously fission, i.e. split into several smaller fragments. During this process two or three neutrons are emitted. These neutrons can induce further fissions if they hit other uranium nucleuses. As the size of a nucleus is very small, neutrons have good chance of escaping the mass of uranium without hitting a nucleus. This chance decreases when the mass is increased, as the probability of hitting a nucleus is proportional to the linear distance a neutron has to travel through uranium to escape. The fraction of neutrons that escape without producing further fissions is analogous to the fraction of the adult males who have no sons in Galton-Watson model. The neutrons produced in a fission induced by a particular neutron are analogous to sons. Critical branching process corresponds to a critical mass. A nuclear explosion is a supercritical branching process.

Theory of branching processes is also useful in studies of chemical chain reactions, cosmic rays,

---

[8] Let $y = f(x)$ and $y_0 = f(x_0)$ where $f'(x_0) \neq 0$, then (see Weisstein, a ):

$$x = x_0 + \sum_{k=1}^{\infty} \frac{(y - y_0)^k}{k!} \left\{ \frac{d^{k-1}}{dx^{k-1}} \left[ \frac{x - x_0}{f(x) - y_0} \right] \right\}_{x = x_0}$$

and population genetics (replace "surname" with "gene"). In this paper we show that it is helpful for understanding scientific citation process. Here the first year citations correspond to sons. Second year citations, which are copies of the first year citations, correspond to grandsons, and so on.

## Appendix B

Let us consider the case when $\lambda \neq \lambda_0$, i.e. a branching process with generating function for the first generation being different from the one for subsequent generations. One can show that the generating function for the total offspring is:

$$\tilde{g}(z) = z f_0(g(z)). \tag{B1}$$

Note that in the case $\lambda = \lambda_0$ we have $f(z) = f_0(z)$ and, because of Eq. A9, $\tilde{g}(z) = g(z)$.

From Eqs. 4 and 5 it follows that $f_0(z) = (f(z))^{\lambda_0/\lambda}$. Substituting this together with Eq. A9 into Eq. B1 we get:

$$\tilde{g}(z) = z \left( \frac{g(z)}{z} \right)^{\frac{\lambda_0}{\lambda}} \tag{B2}$$

This formula may be of some use when the ratio $\frac{\lambda_0}{\lambda}$ is an integer.

As $\frac{\lambda_0}{\lambda} = \frac{\alpha}{1-\alpha} N_{ref}$ and $\alpha \approx 0.1$, $N_{ref} \approx 20$ we have $\frac{\lambda_0}{\lambda} \approx 2$. Eq. B2 reduces to:

$$\tilde{g}(z) = z \left( \frac{g(z)}{z} \right)^2 = z \left( \sum_{n=1}^{\infty} P(n) z^{n-1} \right)^2 = \sum_{n=1}^{\infty} z^n \sum_{l=1}^{n} P(l) P(n-l+1) \tag{B3}$$

where $P(n)$ is given by Eq. 5. The citation probability distribution is thus:

$$\tilde{P}(n) = \sum_{l=1}^{n} P(l) P(n-l+1) \tag{B4}$$

We can easily obtain the large-$n$ asymptotic of $\tilde{P}(n)$, after noticing that only the terms with either $l \ll n$ or $n - l \ll n$ essentially contribute to the sum:

$$\tilde{P}(n) = \sum_{l=1}^{n/2} P(l) P(n-l+1) + \sum_{l=n/2}^{n} P(l) P(n-l+1) \propto$$

$$P(n) \sum_{l=1}^{n/2} P(l) + P(n) \sum_{l=n/2}^{n} P(n-l+1) \propto$$

$$2 P(n) \sum_{l=1}^{\infty} P(l) = 2 P(n)$$

where $P(n)$ is given by Eq. 6. We see that having different first generation offspring probabilities does not change the functional form of the large-$n$ asymptotic, but merely modifies the numerical prefactor.

## Appendix C

Let us start with the self consistency equation for $p_r(\varphi)$, the equilibrium fitness distribution of references:

$$p_r(\varphi) = \alpha \frac{\varphi \times p_p(\varphi)}{\langle \varphi \rangle_p} + (1-\alpha) \frac{\varphi \times p_r(\varphi)}{\langle \varphi \rangle_r} \tag{C1}$$

solution of which is:

$$p_r(\varphi) = \frac{\alpha \times \varphi \times p_p(\varphi) / \langle \varphi \rangle_p}{1 - (1-\alpha) \varphi / \langle \varphi \rangle_r} \tag{C2}$$

One obvious self-consistency condition is that

$$\int p_r(\varphi) d\varphi = 1. \tag{C3}$$

Another is:

$$\int \varphi \times p_r(\varphi) d\varphi = \langle \varphi \rangle_r. \tag{C4}$$

It is easy to see that when the condition of Eq. C3 is satisfied Eq. C4 follows from Eq. C1.

In the case of a uniform distribution of fitness, using Eqs. C2 and C3 we recover Eq. 15.

Let us consider the published papers fitness distribution of the following form:

$$p_p(\varphi) = \begin{cases} 2-a & \text{when} & 0 \le \varphi < 1/2 \\ a & \text{when} & 1/2 \le \varphi < 1 \\ 0 & \text{otherwise} \end{cases} \quad (C5)$$

This distribution reduces to a uniform distribution when $a = 1$. Elementary calculation gives $\langle\varphi\rangle_p = \dfrac{1+a}{4}$ and after substituting Eqs. C5 and C2 into Eq. C3 we get:

$$1 = \frac{4\alpha}{1+a}\left((2-a)\int_0^{1/2}\frac{\varphi d\varphi}{1-\gamma\varphi/\langle\varphi\rangle_r} + a\int_{1/2}^{1}\frac{\varphi d\varphi}{1-\gamma\varphi/\langle\varphi\rangle_r}\right).$$

After integrating we obtain:

$$1 = \frac{4\alpha}{1+a}\frac{\langle\varphi\rangle_r}{\gamma}\times\begin{pmatrix}-1-\dfrac{\langle\varphi\rangle_r}{\gamma}a\ln\left(1-\dfrac{\gamma}{\langle\varphi\rangle_r}\right)-\\ \dfrac{\langle\varphi\rangle_r}{\gamma}2(1-a)\ln\left(1-\dfrac{1}{2}\dfrac{\gamma}{\langle\varphi\rangle_r}\right)\end{pmatrix} \quad (C6)$$

When $\alpha$ is small $\dfrac{\gamma}{\langle\varphi\rangle_r}$ should be very close to 1 and we can replace it with one everywhere in the Eq. C6 except for in $\ln\left(1-\dfrac{\gamma}{\langle\varphi\rangle_r}\right)$. Resulting equation can be easily solved to get:

$$\frac{\gamma}{\langle\varphi\rangle_r} \approx 1 - \exp\left(-\frac{1}{a}\left(\frac{1+a}{4\alpha}+1-2(1-a)\ln(2)\right)\right) \quad (C7)$$

For example, when $\alpha = 0.1$ and $a = 0.2$ we get from Eq. C7 that $\dfrac{\gamma}{\langle\varphi\rangle_r} \approx 1-e^{-14}$. We see that $\dfrac{\gamma}{\langle\varphi\rangle_r}$ is very close to unity, similar to what happened in the case of a uniform distribution of fitness. One can reason that this is true for all fitness distributions which approach a non-zero limit at the maximum value of fitness.

Now we proceed to investigate the fitness distribution, which vanishes at $\phi = 1$:

$$p_p(\varphi) = (\theta+1)(1-\varphi)^\theta \quad (C8)$$

Substituting Eq. C8 into Eq. C2 we get:

$$p_r(\varphi) = \frac{\alpha(\theta+1)(\theta+2)\varphi(1-\varphi)^\theta}{1-\gamma\varphi/\langle\varphi\rangle_r} \quad (C9)$$

After substituting this into C3 and some calculations we arrive at:

$$1 = \frac{\alpha(\theta+1)(\theta+2)}{(\gamma/\langle\varphi\rangle_r)^2}\int_0^1\frac{x^\theta dx}{\dfrac{\langle\varphi\rangle_r}{\gamma}-1+x} - \frac{\alpha(\theta+2)}{\gamma/\langle\varphi\rangle_r} \quad (C10)$$

As acceptable values of $\langle\varphi\rangle_r$ are limited to the interval between $\gamma$ and 1, it is clear that when $\alpha$ is small the equality in Eq. C11 can only be attained when the integral is large. This requires $\dfrac{\langle\varphi\rangle_r}{\gamma}$ being close to 1. And this will only help if $\theta$ is small. In such case the integral in C11 can be approximated as $\displaystyle\int_{\frac{\langle\varphi\rangle_r}{\gamma}-1}^{1}\frac{x^\theta dx}{x} = \frac{1}{\theta}\left(1-\left(\frac{\langle\varphi\rangle_r}{\gamma}-1\right)^\theta\right)$.

Substituting this into Eq. C11 and replacing in the rest of it $\dfrac{\langle\varphi\rangle_r}{\gamma}$ with unity we can solve the resulting equation to get:

$$\frac{\langle\varphi\rangle_r}{\gamma}-1 \approx \left(\frac{\alpha-\dfrac{\theta}{\theta+2}}{\alpha(\theta+1)}\right)^{\frac{1}{\theta}} \quad (C11)$$

For example, when $\alpha = 0.1$ and $\theta = 0.1$ we get from Eq. C11 that $\dfrac{\langle\varphi\rangle_r}{\gamma}-1 \approx 6\times 10^{-4}$.

Note that Eq. C11 gives a real solution only when $\alpha > \dfrac{\theta}{\theta+2}$ or

$$\theta < \frac{2 \times \alpha}{1-\alpha}. \qquad (C12)$$

If $\theta$ is too big and the condition of Eq. C12 is violated there is no stationary distribution which can satisfy Eq. C1. The distribution is forever changing without reaching a stationary state. Numerical simulations indicate that $\langle \varphi \rangle_r$ is growing from year-to year, but always remains less than $\gamma$. This means that top-fit papers are supercritical (see Eq. 11), the fraction of supercritical papers decreases from year-to year, but never vanishes entirely.